\documentclass[aps, prb, showpacs, preprint, amsmath, letterpaper]{revtex4-1}

\usepackage{graphics}
\usepackage{graphicx}
\usepackage{amssymb}
\usepackage{bm}

\begin{document}

\title{Symmetry of magnetic correlations in spin-triplet superconductor UTe$_2$}

\author{Nicholas P. Butch$^{1,2,*}$, Sheng Ran$^{1,2,3}$, Shanta R. Saha$^{1,2}$, Paul M. Neves$^{1,2}$, Mark P. Zic$^{1,2}$, Johnpierre Paglione$^{1,2}$, Sergiy Gladchenko$^{1}$, Qiang Ye$^{1,4}$, Jose A. Rodriguez$^{1}$}

\affiliation{$^1$ NIST Center for Neutron Research, National Institute of Standards and Technology, Gaithersburg, MD 20899, USA
\\$^2$ Maryland Quantum Materials Center, Department of Physics, University of Maryland, College Park, MD 20742, USA
\\$^3$ Department of Physics, Washington University in St. Louis, St. Louis, MO 63130, USA  
\\$^4$ Department of Materials Science and Engineering, University of Maryland, College Park, MD 20742-2115, USA
\\$*$ email: nbutch@umd.edu
}

\date{\today}

\begin{abstract}

The temperature dependence of the low-energy magnetic excitations in the spin-triplet superconductor UTe$_2$ was measured via inelastic neutron scattering in the normal and superconducting states. The imaginary part of the dynamic susceptibility follows the behavior of interband correlations in a hybridized Kondo lattice with an appropriate characteristic energy. These excitations are a lower-dimensional analogue of phenomena observed in other Kondo lattice materials, such that their presence is not necessarily due to dominance of ferromagnetic or antiferromagnetic correlations. The onset of superconductivity alters the magnetic excitations noticeably on the same energy scales, suggesting that these changes originate from additional electronic structure modification.

\end{abstract}
\maketitle

Spin-triplet superconductivity was recently discovered in UTe$_2$ \cite{ran_nearly_2019}. The superconductivity is characterized by large and anisotropic upper critical fields that all exceed the paramagnetic limit, pointing to unconventional spin-triplet pairing 
\cite{ran_nearly_2019,aoki_unconventional_2019}. Superconductivity is limited by a magnetic phase transition at 35~T\cite{ran_extreme_2019,knebel_field-reentrant_2019}, and the field-polarized state contains another reentrant superconducting phase above 40~T\cite{ran_extreme_2019}. Superconductivity in UTe$_2$ is believed to be topologically nontrivial because of observations of chiral in-gap surface states \cite{jiao_chiral_2020}, a double transition in specific heat, and broken time reversal symmetry as detected by optical Kerr rotation \cite{hayes_multicomponent_2021}, which suggest the presence of a complex, two-component superconducting order parameter. Consistent with a p-wave orbital symmetry, the superconducting gap is nodal \cite{bae_anomalous_2021,metz_point-node_2019,ishihara_chiral_2021}.

Superconductivity emerges from a renormalized electronic structure of hybridized f-electrons. UTe$_2$ exhibits archetypal heavy fermion features, namely a large low-temperature specific heat and local maxima in temperature-dependent electrical resistivity and magnetic susceptibility below room temperature \cite{ran_nearly_2019}, a Kondo hybridization gap in scanning tunneling spectroscopy \cite{jiao_chiral_2020}, quadratic temperature dependence of low-temperature resistivity \cite{eo_anomalous_2021}, linear temperature dependence of thermoelectric power\cite{niu_fermi-surface_2020}, and a Drude peak in optical conductvity \cite{mekonen_optical_2021}. Angle resolved photoemission (ARPES) measurements show that the band structure of UTe$_2$ is dominated by two intersecting one-dimensional sheets \cite{miao_low_2020}. The heavy electron states result from hybridization between these highly-dispersive bands with f-electron states near the chemical potential, as suggested by dynamical mean field theory (DMFT) calculations \cite{miao_low_2020,mekonen_optical_2021}, while ARPES also reveals an additional Fermi pocket that is three-dimensional and potentially heavy. 

Magnetic interactions play an important role in UTe$_2$. Low-temperature neutron diffraction demonstrates the lack of long range magnetic order in the normal state\cite{hutanu_low-temperature_2020}. Yet several measurements suggest proximity to a ferromagnetic instability. A scaling analysis of the magnetization suggests the influence of ferromagnetic quantum critical fluctuations \cite{ran_nearly_2019}. Muon spin relaxation measurements have shown that significant spin fluctuations in the normal state strengthen upon cooling into the superconducting state \cite{sundar_coexistence_2019}. Nuclear magnetic resonance measurements are difficult at low temperature in the normal state due to long relaxation times \cite{tokunaga_125te-nmr_2019}. Optical Kerr rotation experiments reveal that vortices consist of magnetically polarizable normal cores \cite{wei_interplay_2021}.

Recent inelastic neutron scattering experiments complicate the picture, as they have not found obvious signatures of ferromagnetic fluctuations, but rather magnetic excitations at incommensurate wavevectors. These might originate in RKKY interactions or Fermi surface nesting \cite{duan_incommensurate_2020}, or rod-like excitations due to spin-ladder interactions \cite{knafo_low-dimensional_2021}. Intriguingly, a change in the inelastic neutron scattering emerges in the superconducting state, near energies of 1~meV \cite{duan_resonance_2021,raymond_feedback_2021}, which is suggested to be a new type of superconducting resonance. Recent calculations show that a dominant antiferromagnetic susceptibility can be consistent with spin-triplet superconductivity \cite{kreisel_spin_2021}.

We performed a series of inelastic neutron scattering experiments in the crystallographic a-b plane that demonstrate the detailed energy-dependence and anisotropy of the magnetic excitations. The excitations evolve over a broad range of temperatures, from the weakly correlated high-temperature state into the superconducting state below 1.6~K. These excitations are signatures of the heavy electron band structure, similar to several other Kondo lattice systems, but with a lower dimensionality due to the UTe$_2$ structure. They do not a priori imply a tendency toward a specific type of long-range magnetic order. These measurements also show that the change in the excitation spectrum in the superconducting state is a phenomenon that occurs over energies of several meV, suggesting a substantial change in magnetic correlations at low temperatures.

\begin{figure}
\includegraphics[angle=0,width=150mm]{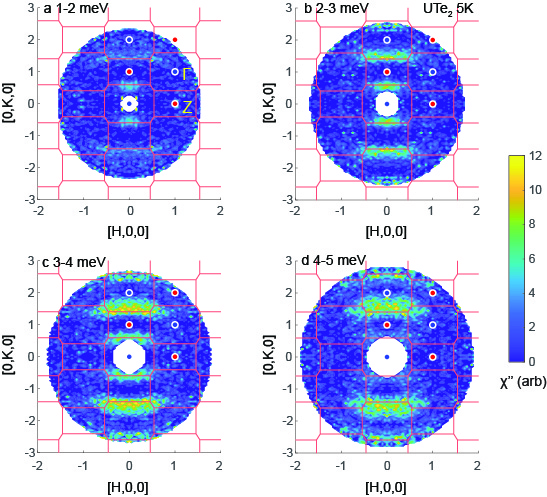}
\caption{Imaginary part of the dynamic magnetic susceptibility $\chi''(Q,E)$ of UTe$_2$ at 5~K, in the regime of strong hybridization, as measured by neutron scattering. The susceptibility in the [H,K,0]~plane, corresponding to the crystallographic a-b plane is strongest in the energy intervals a) 1-2~meV, b) 2-3~meV, c) 3-4~meV, d) 4-5~meV. These inelastic magnetic excitations closely follow the Brillouin zone edges (red) along the K direction. These excitations do not extend much beyond $H=0.5$ and are absent in symmetrically equivalent Brillouin zones that are not centered on $H=0$. Blue and red circles mark the Brillouin zone centers $\Gamma$ and Brillouin zone faces $Z$, respectively, which are neighbors in the [H,K,0]~plane due to the body-centered crystal structure. }
\label{Fig1}
\end{figure}

The scattered neutron intensity $S$, measured as a function of transfer of momentum $Q$ and energy $E$, is proportional to a temperature factor times the imaginary part of the dynamic magnetic susceptibility $\chi''$, by $\chi''(Q,E) = (1-{e^\frac{-E}{k_{B}T}}) S(Q,E)$, where $k_B$ is the Boltzmann constant and $T$ is the temperature. In UTe$_2$, the dominant feature in $\chi''$ is a well-defined $Q$-dependence with peak intensity at $E=4$~meV. As shown in Fig.~\ref{Fig1}, at 5~K these magnetic excitations essentially form stripes long in the $H$ direction, the crystallographic a-axis, and modulated along the $K$ direction, the b-axis. As the Brillouin zone (BZ) overlays in Fig.~\ref{Fig1} make clear, these excitations are confined to the edges of the BZs, occurring at values of $K = 0.6 = \frac{1}{2}(1+\frac{b^2}{c^2})$, for lattice constants $b=6.08~\mathrm{\AA}$ and $c=13.8~\mathrm{\AA}$, and symmetrically equivalent values such as $K=1.4$ and $K=2.6$. However, the excitations are constrained to near $H=0$ and are not detected at the edges of any BZ's that do not fall on the $K$ axis, even though they are symmetrically equivalent. Weak excitations at larger $H$ values \cite{duan_resonance_2021} are not evident here. Overall, similar excitations extend along the c-direction \cite{duan_incommensurate_2020,knafo_low-dimensional_2021}, so we conclude that these excitations are observed only for $Q$ perpendicular to the bulk magnetic easy axis.

\begin{figure}
\includegraphics[angle=0,width=150mm]{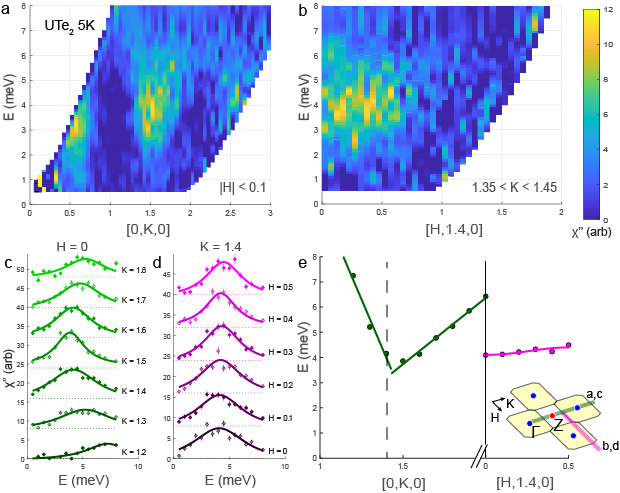}
\caption{The anisotropic $Q$-dependence of the magnetic susceptibility $\chi''(Q,E)$ in UTe$_2$ at 5~K along different slices of the [H,K,0] plane. a) $\chi''(Q,E)$ as a function of K at constant H = 0. Along this direction, the dispersion is V-shaped and sharp, with energy minima at K = 0.6, 1.4 and 2.6, corresponding to the Brillouin zone boundaries. b) $\chi''(Q,E)$ as a function of H at constant K = 1.4. The susceptibility is flat in the H direction, extending to $H=0.5$, the edge of the Brillouin zone. c,d) Effective constant-$Q$ scans show the $E$-dependence of $\chi''$ at selected values of $K$ from a) and $H$ from b). Curves are fits to Lorentzian lineshapes, which show that the width of the magnetic excitations is of order the peak energy. Whereas the lineshapes are rather dependent of $K$, there is relatively little $H$-dependence until the excitation disappears at the edge of the Brillouin zone. These fits are used to calculate the magnetic dispersion in e) along these two directions. A clear anistropy appears about the BZ edge (dashed line). The inset shows a $Q$ map of the data slices.}
\label{Fig2}
\end{figure}

Looking more closely at the $Q$-dependence, it is clear that the excitations disperse asymmetrically. Figure \ref{Fig2} shows contour plots of $\chi''$ at 5~K, comparing the magnetic dispersion along a) the b-axis and b) the a-axis at the BZ edge. To more easily see the $Q$-dependence in the vicinity of $[0,1.4,0]$, these data are broken up into constant-$Q$ scans in Fig.~\ref{Fig2}c. Fits of Lorentzian lineshapes can describe the data adequately. The energy of peak intensity can be tracked as a function of $Q$ (Fig.~\ref{Fig2}d). There is an obvious, sharp minimum in the peak position at $K = 1.4$ with a value of 4~meV, which remains approximately constant as a function of H along $[H,1.4,0]$. Thus the excitation is not dispersive along the BZ edge, but only along the perpendicular direction. Note also that the dispersion along $[0,K,0]$ is very asymmetric about the minimum at $K=1.4$, with a steeper slope of -16~meV~$\mathrm{\AA}$ at smaller $K$ and shallower slope of 5.5 meV~$\mathrm{\AA}$ at larger $K$. This follows the asymmetry of the paramagnetic body-centered unit cell - note that the excitations for $K<1.4$ actually extend along a BZ face, while for $K>1.4$ they are inside the volume of the BZ as they approach the BZ center $\Gamma$. 

Another outstanding feature is the width in $E$ of the excitatations in UTe$_2$, which is  comparable to the peak excitation energy, even at $Q$ values where the excitations are sharpest. Despite these large widths, which imply substantially shortened excitation lifetimes or a distribution of transitions, the excitations are clearly peaked at nonzero $E$ - therefore, these are inelastic features that are separated from the ground state by a finite energy gap. The peak energy of 4~meV matches well the hybridization gap determined in scanning tunneling spectroscopy measurements \cite{jiao_chiral_2020}, suggesting a connection to the electronic structure. 

\begin{figure}
\includegraphics[angle=0,width=150mm]{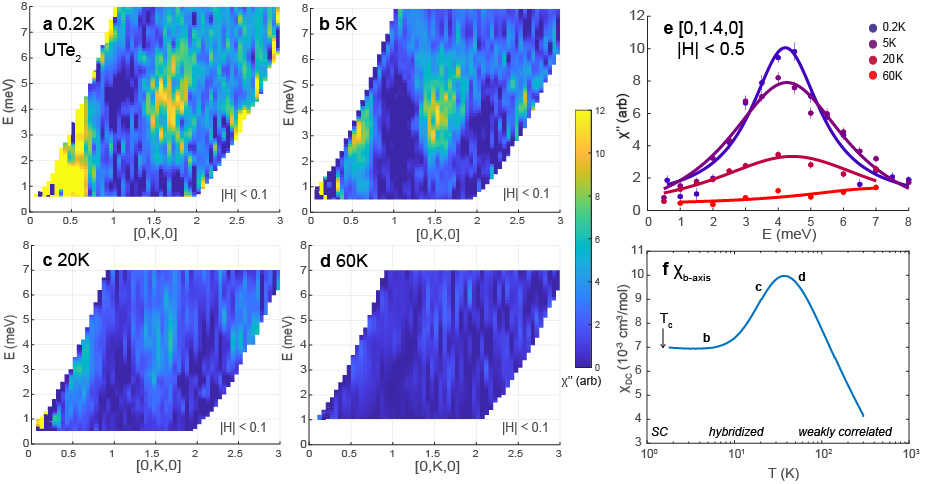}
\caption{Temperature dependence of the magnetic susceptibility in UTe$_2$. $\chi''(Q,E)$ as a function of K at the BZ edge at a) 0.2~K in the superconducting state, at b) 5~K in the strongly hybridized, renormalized Fermi liquid state, at c) 20~K where hybridization is weaker, and d) at 60~K where correlations between local f electrons and light bands are weakest. e) As temperature increases, the total spectral weight decreases but the magnetic susceptibility remains peaked at the same $q$ value until it is no longer detectable above background. f) This behavior follows closely the temperature-dependence of the bulk magnetic susceptibility $\chi$ along the b-axis (K direction)\cite{ran_nearly_2019}, which shows a hump structure characteristic of heavy fermion Kondo lattices (1~cm$^3$/mol = 4$\pi \times 10^{-6}$ m$^3$/mol). The letters indicate the temperatures of the corresponding neutron data.}
\label{Fig3}
\end{figure}

The temperature dependence of the excitations at $K=1.4$ is shown in Fig.~\ref{Fig3}. Cooling below 5~K into the superconducting state yields an excitation spectrum that appears to remain mostly the same. In contrast, on warming from 5~K to 20~K, the intensity decreases and the peak position increases slightly, but the excitations maintain their $Q$-dependence and energy gap. Importantly, they do not move to lower $E$ and become quasielastic, as might occur to magnetic correlations at temperatures above a magnetic phase transition. However, by 60~K the excitations are not discernable over the background. This temperature trend follows closely the low-field magnetic susceptibility along the crystallographic b-axis, which is a signature of the low-temperature development of the renormalized heavy fermion or hybridized electron state in UTe$_2$.

Very similar inelastic BZ edge magnetic excitations are seen in the paramagnetic state in URu$_2$Si$_2$ \cite{butch_symmetry_2015}. Although the ground states are very different, URu$_2$Si$_2$ and UTe$_2$ share several important features. Both uranium compounds exhibit heavy fermion behavior, namely peaks as a function of temperature in the electrical resistivity and magnetic susceptibility, signifying similar energy scales. Also, they both have body-centered crystal structures, highlighting a key geometric difference - whereas URu$_2$Si$_2$ is tetragonal and exhibits inelastic magnetic excitations following the square BZ edges, only lines of excitations along one edge are observed in orthorhombic UTe$_2$. The $E-Q$ dispersions are further similar in that the slopes are shallow along the BZ edge and much sharper perpendicular to the edge. Although the magnetic excitations in URu$_2$Si$_2$ fall at incommensurate $Q$ vectors, they arise from interband scattering \cite{brandow_finite-temperature_1988}, as has been observed in other Kondo lattice compounds \cite{goremychkin_coherent_2018}. Most importantly, these incommensurate excitations persist in URu$_2$Si$_2$ regardless of whether the ordered state is hidden order, antiferromagnetism \cite{butch_distinct_2016}, or ferromagnetism \cite{butch_ungapped_2020}. 

Therefore, the presence of incommensurate excitations in UTe$_2$ is insufficient to draw conclusions about incipient static magnetic order. Calculations do not provide strong constraints, as the dominant magnetic interactions in UTe$_2$ are easily tuned from ferromagnetic to antiferrogmagnetic \cite{ishizuka_periodic_2021}. Many experiments suggest ferromagnetic interactions: muon spin relaxation \cite{sundar_coexistence_2019} and optical Kerr rotation experiments \cite{hayes_multicomponent_2021,wei_interplay_2021} are consistent with low-temperature ferromagnetic correlations. As noted previously \cite{duan_incommensurate_2020,knafo_low-dimensional_2021} the absence of clear neutron scattering intensity near BZ centers as is seen in UCoGe \cite{stock_anisotropic_2011} suggests that any magnetic fluctuations in UTe$_2$ that might be associated with an incipient order parameter or quantum criticality are weak, but this is consistent with the bulk magnetization. 

To account for this hybridization, it will be necessary to carry out further electronic structure and magnetic susceptibility calculations with higher energy resolution. Available calculations suggest that the strongest atomic exchange interaction in UTe$_2$ is ferromagnetic, between uranium dimers, with antiferromagnetic correlations parallel to the chains along the a-axis \cite{miao_low_2020,shishidou_topological_2021}, but this does not readily explain the measured $\chi''(Q,E)$. The observed b-axis modulation has been considered to arise from the electronic structure, either Fermi surface nesting or RKKY exchange \cite{duan_incommensurate_2020}. However, calculations to date have not yet addressed an important experimental point, namely that the measured $\chi''(Q,E)$ is peaked at small but finite energy, not at zero energy, the latter condition relevant to static magnetic order. Generally, given that these excitations appear to follow the temperature dependence of the b-axis bulk magnetic susceptibility, we expect that they will be quite robust as a function of magnetic field and could play a role in the magnetic transition at 35~T, and by extension, in the magnetically ordered phase above 1.5~GPa \cite{lin_tuning_2020,ran_expansion_2021,valiska_magnetic_2021}. 

\begin{figure}
\includegraphics[angle=0,width=160mm]{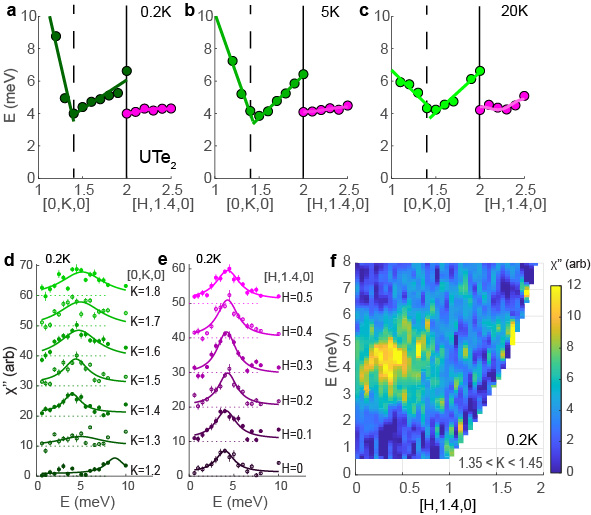}
\caption{Temperature dependence of the magnetic dispersion near the BZ edge at $[0,1.4,0]$. The lower-$K$ dispersion is steeper in a) the superconducting state, and gradually softens in b) the normal state and at c) higher temperatures. d,e) Effective constant-$Q$ scans show the $E$-dependence of $\chi''$ at selected values of $K$ from and $H$ from a). Note the relative intensity decreases at $[0,1.4,0]$ compared to 5~K (Fig.~\ref{Fig2}c). f) $\chi''(q,E)$ as a function of H at constant $K = 1.4$ in the superconducting state. Compare to Fig.~\ref{Fig2}b.}
\label{Fig4}
\end{figure}

The situation at lowest temperatures, in the superconducting state, brings an interesting twist. As Fig.~\ref{Fig3} shows, the magnetic excitation spectrum is similar at 0.2~K and 5~K. Indeed, a significant difference is not expected at these energies, given the 1.6~K critical temperature and 0.25~meV gap observed in STM \cite{jiao_chiral_2020}. Therefore, the recently reported feature at 1~meV in the superconducting state\cite{duan_resonance_2021,raymond_feedback_2021} suggests either a very large energy excitation of the superconducting state, or a modification of the established low-energy spin-fluctuations. In our measurements, it is not possible to conclusively identify the 1~meV feature at $K=0.6$ because of the high background in that $Q$ range, but it is not resolved at $K=1.4$. Yet this feature is connected to the broader magnetic excitation and other changes in the magnetic spectrum are observed at 0.2~K. First, the dispersion along $[0,K,0]$ is steeper for $K<1.4$ and slightly gentler for $K>1.4$ (Fig.~\ref{Fig4}a-d) than at 5~K. In fact, the dispersion is weakly temperature-dependent at least to 20~K. Second, a slight decrease in intensity is observed near $[0,1.4,0]$ in the superconducting state (Fig.~\ref{Fig4}e,f), when compared to the data at 5~K (Fig.~\ref{Fig2}).

This behavior suggests that in the superconducting state, $\chi''$ changes on energy scales even larger than 1~meV, and that previously reported feature reflects a larger change in the magnetic excitation spectrum. Such a broad change is difficult to reconcile with an interpretation in terms of a superconducting spin resonance. Since these magnetic excitations have their origin in the heavy fermion band structure, low-temperature changes in the electronic structure are likely responsible, and this is evidence that the superconducting state involves heavy quasiparticles. The change in the magnetic excitations may additionally correlate with low-temperature changes in the magnetic response in muon spin relaxation \cite{sundar_coexistence_2019} and nuclear magnetic resonance \cite{nakamine_superconducting_2019} inside the superconducting state.

In summary, inelastic neutron scattering measurements reveal magnetic excitations that are consistent with Kondo lattice phenomena. Specifically, the excitations follow BZ edges, obey the paramagnetic structural symmetry, and the temperature evolution of the heavy fermion bulk magnetic susceptibility along the b-axis. By analogy with other Kondo lattice compounds, these excitations are not directly related to an incipient magnetic ordered state. In the superconducting state, the magnetic excitations near $[0,1.4,0]$ decrease in energy, likely related to a change in the electronic structure.

\section{methods}
Single crystals of UTe$_{2}$ were synthesized by the chemical vapor transport method using iodine as the transport agent \cite{ran_comparison_2021}. The crystals are from synthesis batches that have been previously characterized \cite{ran_nearly_2019,ran_extreme_2019} and exhibit consistent properties.

Crystal orientation was determined by Laue x-ray diffraction performed with a Photonic Science x-ray measurement system. 1.2~g of single crystals, ranging in mass from 0.01~g to 0.1~g were coaligned and affixed to two copper plates using CYTOP fluoropolymer and Fomblin fluorinated grease. Inelastic neutron scattering experiments were performed on the MACS spectrometer and preliminary measurements on the DCS spectrometer at the NIST Center for Neutron Research. Background subtraction and symmetrization are elaborated upon in the supplement.

\section{Data Availability}

The data that support the results presented in this paper and other findings of this study are available from the corresponding author upon reasonable request.

\section{Acknowledgments}
We thank Gabi Kotliar and Daniel Agterberg for helpful discussions and Yegor Vekhov for experimental assistance. Access to MACS and support for PMN and MPZ were provided by the Center for High Resolution Neutron Scattering, a partnership between the National Institute of Standards and Technology and the National Science Foundation under Agreement No. DMR-2010792. Research at the University of Maryland was supported by the National Institute of Standards and Technology, the Department of Energy Award No. DE-SC-0019154 (sample characterization) and the Gordon and Betty Moore Foundation's EPiQS Initiative through Grant No. GBMF9071 (materials synthesis). Identification of commercial equipment does not imply recommendation or endorsement by NIST. Error bars throughout this manuscript correspond to an uncertainty of one standard deviation.

\section*{Author contributions}
NPB conceived and designed the study. SR and SRS synthesized the single crystals. SR, SRS, PMN, and MPZ coaligned the crystals. NPB, PMN, MPZ, SG, QY, and JAR performed the neutron scattering measurements. NPB analyzed the data and wrote the manuscript, with contributions from all authors.

\section{Competing Interests}
The authors declare no competing interests.

\clearpage

\end{document}